\documentclass{aa}  

\usepackage{txfonts}
\usepackage{graphicx}
\usepackage[normalem]{ulem}
\usepackage{amsmath}
\usepackage{comment}
\usepackage{natbib}
\bibpunct{(}{)}{;}{a}{}{,} % to follow the A&A style
\usepackage{xcolor}

\usepackage{pifont}

\begin{document} 

\title{Update of the CODE catalogue\\
and some aspects of the dynamical status of Oort Cloud comets.}

\titlerunning{CODE catalogue update}

\authorrunning{P.A.Dybczyński \& M.Królikowska}

\author{Piotr A. Dybczyński\inst{\ref{inst1}} and  Małgorzata Królikowska\inst{\ref{inst2}} }

\institute{
	 Astronomical Observatory Institute, Faculty of Physics and Astronomy, A.Mickiewicz University, Słoneczna 36, 60-286 Poznań, Poland, \email{dybol@amu.edu.pl}\label{inst1} \and Centrum Badań Kosmicznych Polskiej Akademii Nauk (CBK PAN), Bartycka 18A, Warszawa, Poland, \email{mkr@cbk.waw.pl} \label{inst2}
}

\date{Received XXXXXX; accepted XXXXX}

  \abstract {The outer Solar System is believed to host a vast reservoir of long-period comets (LPCs), but our understanding of their spatial distribution and dynamical history remains limited due to observational biases and uncertainties in orbital solutions for really observed comets.}
  {We aim to provide a comprehensive and dynamically homogeneous orbital database of LPCs to support the study of their origin, evolution, dynamical status, and 6D distribution of orbital elements.}     
  { We updated the Catalogue of Cometary Orbits and their Dynamical Evolution (CODE catalogue) by computing original and future barycentric orbits, orbital parameters at previous and next perihelion, using full Monte Carlo swarms of real comets for the uncertainty estimation and taking into account the planetary, Galactic and passing stars perturbations according to the latest data and algorithms.}   
{This update of  the CODE catalogue focuses on the dynamical status of near-parabolic comets. Using current stellar data, we formulated new constraints for dynamically new comets. Now, the CODE database includes 983 orbital solutions for 369 comets with full uncertainty estimates and dynamical classifications, covering nearly all comets with original semi-major axes exceeding 10,000\,au and discovered before 2022, as well as all LPCs discovered beyond 10\,au from the Sun  during this period, and over 80\% of  the known LPCs with perihelion distances beyond 7\,au.} {}
   

   \maketitle

\section{Introduction}\label{intro}

Three-quarters of a century have passed since \cite{oort:1950} proposed that the Solar System is surrounded by a distant, nearly spherical cloud of comets, now known as the Oort Cloud. According to this hypothesis, only a tiny fraction of these comets are occasionally perturbed towards the inner Solar System and become observable. Although the number of known Oort spike comets (i.e. those with original semi-major axes larger than 10,000 au) has grown significantly since Oort’s time -- by more than a factor of 20 -- our ability to characterise their true spatial distribution as a function of the heliocentric distance remains severely limited. This limitation arises not only from observational selection effects, but also from the need for precise orbital solutions and a reliable assessment of their dynamical histories. Therefore, the best possible cometary orbits must be continuously collected.

On 7 July 2008, the last (seventeenth) edition of the Catalogue of Cometary Orbits was announced in IAU Circular No. 8958 \citep{IAUC-8958}. The IAU Minor Planet Center \citep[][MPC]{MPC_DB_Search} continues to calculate and publish cometary orbits. To our knowledge, there are only two other massive alternative sources:  the JPL Small-Body Database Browser \citep{JPL_SBDB} and \cite{Nakano_Notes}.

In all three places,  the orbits of both short- and long-period comets are presented. At the JPL, they only offer the osculating orbit. MPC and Nakano also calculate the original and future semi-major axis reciprocal $1/a$. Only Nakano presents individual residuals for the observations used in the orbit determination. However, when searching for the source of long-period comets and asking what their dynamical age is, one needs a full set of the original orbital elements that allow propagation  them to the past. 
When we introduced the \textit{Catalogue of Cometary Orbits and their Dynamical Evolution} (CODE catalogue)\footnote{\tiny{\tt https://apollo.astro.amu.edu.pl/CODE}} \footnote{\tiny{\tt https://code.cbk.waw.pl/}} in \cite{kroli-dyb:2020}, we restricted ourselves to  the near-parabolic comets only, but we decided to offer much more information suitable for  the LPC source region and origin studies.

The CODE catalogue provides a curated and dynamically consistent set of orbital solutions for near-parabolic comets, including those with varying levels of orbital precision. Each entry is accompanied by a quality assessment to support further analysis. For every comet, the catalogue includes both original (pre-entry) and future (post-planetary encounter) orbital elements, as well as orbits propagated to the previous and next perihelion passages,  termed 'previous' and 'next' orbits. All orbital solutions are supplemented with comprehensive uncertainty estimates derived from dedicated numerical integrations of large virtual comet (VC) swarms.  This VC~approach, introduced by \cite{sitarski:1998} and next applied in detail to LPCs in \cite{krolikowska:2001, krolikowska:2004, krolikowska:2006a}, and \cite{kroli-dyb:2010} and later, uses Monte Carlo–generated sets of orbital elements to propagate observational uncertainties in dynamical studies.
%This approach is also used in the CODE~Catalogue for orbital element estimates during each step of the orbital evolution.}

In addition, each comet is assigned a dynamical status based on its origin and its evolution  towards the previous perihelion. The CODE catalogue is thus intended to serve as a robust foundation for studying the dynamical structure, source regions, and long-term evolution of the outer Solar System comet population. 

The presented update represents a significant enhancement of the CODE catalogue in terms of the number of comets included and the richer functionality. The catalogue itself was thoroughly described in \cite{kroli-dyb:2020}, with a later update briefly announced in \cite{Kroli-Dyb:2023}. Here, the current strengths of the catalogue are illustrated, among other ways, through two samples of LPCs: comets with a large perihelion distance and comets discovered at a large heliocentric distance. Their activity has been observed more frequently in recent years as a result of growing capabilities in detecting increasingly faint objects. Moreover, with the upcoming operation of the Vera C.~Rubin Observatory (see \citet{Inno:2024, Juric:2023}, and \cite{Ivezic:2019} for an overview of the LSST capabilities), we expect a large-scale discovery of comets at substantial heliocentric distances.

A detailed description of the new elements in the CODE catalogue is provided in Sect.~\ref{sec:CODEnews}. 
Sect.~\ref{sec:CODE_a-distributions} presents a comprehensive analysis of the distribution of original semi-major axes, taking into account not only the nominal orbits, but also the uncertainties in orbital determinations. Next, in Section \ref{prev_next}, we describe the 'previous' and 'next' orbits, i.e. the orbits propagated to the previous and next perihelion passage. We also address the problem of accessing the dynamical status of all LPCs.
In Sect.~\ref{sec:distant_discoveries}, we discuss a sample of comets discovered at heliocentric distances beyond 10\,au, with particular attention to how well their dynamical status is currently known.   Sect.~\ref{sec:CODE_large_q} discusses the subset of comets with perihelion distances greater than 7\,au, based on both the CODE catalogue and the JPL Small-Body Database.

\begin{table}[ht]
    \caption{Characteristics of LPC sample in the CODE catalogue using {\it preferred} orbits.} 
    \label{tab:CODE_in_brief}
\setlength{\tabcolsep}{9.0pt}    \centering
    \begin{tabular}{ccc}
	\hline \hline 
    Range of $1/a_{\rm ori}$ & Number &  Completness \\
    ~[in au$_{-6}$ units]  & of LPCs    &          \\    
	\hline  
%    < 0 & 18 & + \\
%    ~~~~0 -- 100    &   277                &  +      \\
    < 100 & 295 & + \\
    100 -- 200  &    36                &  almost  \\
    200 -- 400  &    27                &  $\sim$ 50\%  \\
    >400        &    11                &  poor   \\
	\hline  
    \end{tabular}
\end{table}

\section{What is new in the updated CODE catalogue}\label{sec:CODEnews}

The current update is two-pronged. First, orbital solutions for 57 new LPCs have been added; the majority of them have original semi-major axes exceeding 10,000 au. This group includes the following:
%{\tt https://minorplanetcenter.net/}
\begin{itemize}
\item  The complete sample of 24 LPCs discovered in the years 2013–2017 and having $q < 3.1$\,au; data taken from the Minor Planet Center (MPC) database in January-February 2025\footnote{Usually, the observation sets for orbit determination we retrieve from the MPC database \citep{MPC_DB_Search}, except for rare cases when we supplement these sets with other observations. Since the observations sets of observable comets are continuously supplemented with new measurements in the MPC database, we also provide the time intervals when data were retrieved.};
\item All LPCs discovered in 2021 except C/2021~A6 (PanSTARRS), which has already been included in the previous version of the catalogue. These are 13 LPCs from the Oort spike and four with $ 100 < 1/a_{\rm ori} < 300$\,au$_{-6}$   \footnote{Following \cite{Kroli-Dones:2023}, we use the notation au$_{-6}$ to denote $10^{-6}$\,au as the unit used in describing a semi-major axis reciprocal.}; positional data retrieved from the MPC database in January-March 2025;
\item Seven LPCs discovered in 2022, including two LPCs with $100 < 1/a_{\rm ori} < 300$\,au$_{-6}$; data retrieved from the MPC database in March 2025;
\item C/2025 D1 (Groeller) as having  the largest $q$ among currently known LPCs ($q=14.1$\,au). This comet will pass its perihelion in May~2028; data taken from MPC in March~2025;
\item Eight other comets previously not included in the CODE database for different reasons. For C/1958~R1 (Burnham-Slaughter) and C/1959~X1 (Mrkos) we found additional observations announced by \cite{vanBies:1961, vanBies:1962, vanBies:1966} and \cite{Kresak-Antal:1966}. The third comet with measurements collected from the literature is C/1962~C1 (Seki-Line) (for details see its description in the CODE catalogue). The remaining five include C/2014~W10 (PANSTARRS) with a very poor previously known orbit, C/2020~H5 (Robinson) still observed when the previous update took place and simply missed by us: C/2004~YJ35 (LINEAR), C/2015~K7 (COIAS), and C/2019~N1 (ATLAS).
\end{itemize}

The second component of the update involves revisions to  the previously included comets. We updated  the orbital solutions for 20 LPCs discovered between 2016 and 2020, as well as for C/2021~A6. Additional observations for them were retrieved from the MPC database in the autumn of 2024. Some Oort spike comets discovered  during that period still remain under observation.

\begin{figure}
    \centering
    \includegraphics[width=1.00\linewidth]{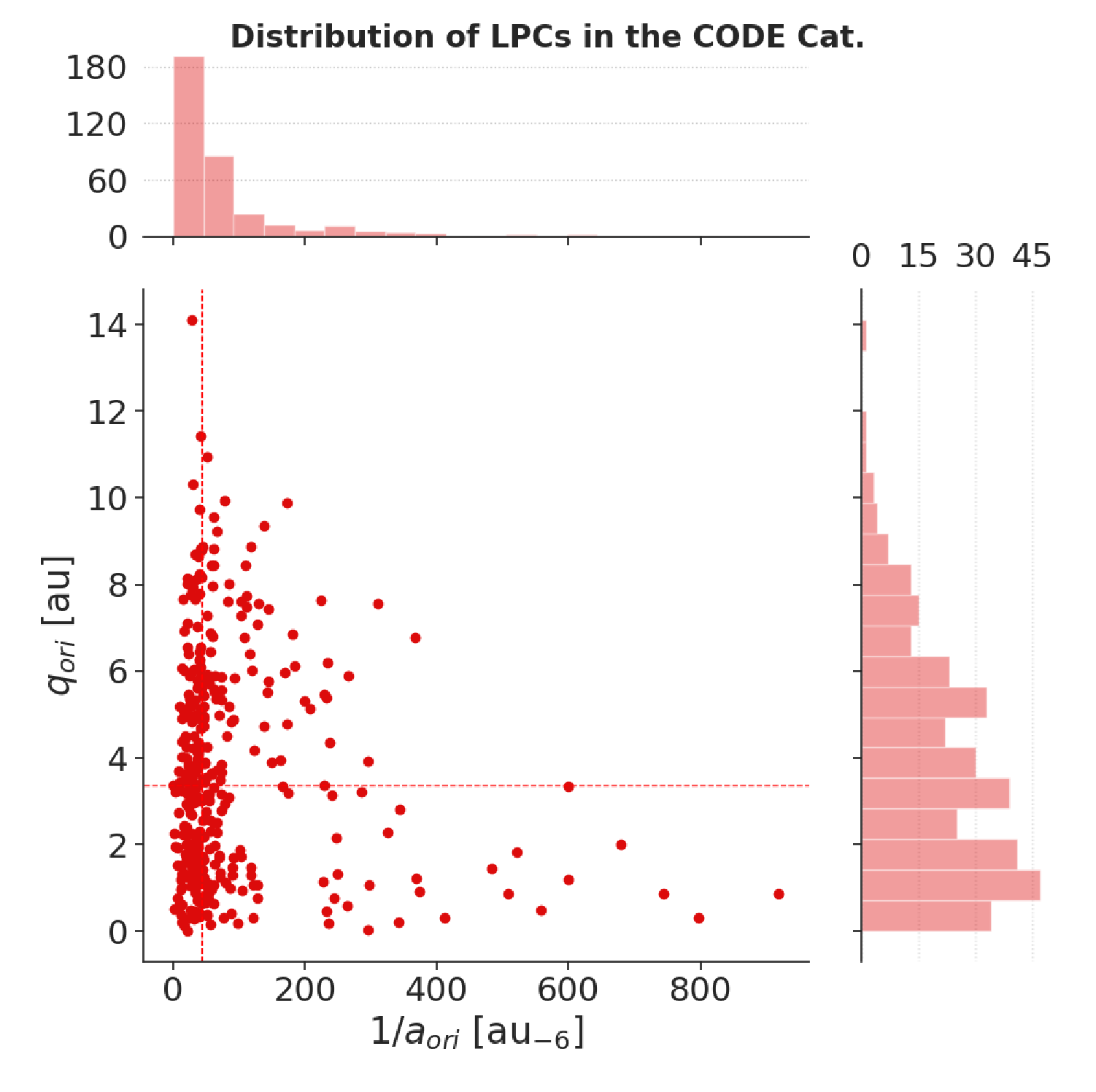}
    \caption{ Distribution $1/a_{\rm ori}$ vs original perihelion distance for the sample of all LPCs in the CODE catalogue, excluding 17 LPCs with hyperbolic original orbits; Shown are parameters for orbital solutions marked as {\it preferred for studies of the past motion}. Median values of $1/a_{\rm ori}$ and $q_{\rm ori}$ are shown using dotted red lines. 
    }
    \label{fig:CODE_q_vs_aori}
\end{figure}

These activities have led to two important enhancements to the catalogue. First, the CODE catalogue now appears to include a complete sample of LPCs discovered between 1900 and 2021 with original semi-major axes greater than 10, 000\,au, and a nearly complete sample with $100 < 1/a_{\rm ori} < 300$\,au$_{-6}$ (i.e., $3,400 < a_{\rm ori} < 10,000$\,au). Second, the catalogue now includes 80\% of all known LPCs with perihelion distances greater than 7\,au. As of March 2025, 56 such objects were known, of which only 12 are not included in the CODE catalogue; see Sect.~\ref{sec:CODE_large_q}. 
In summary, the CODE database now contains 983 orbital solutions for 369 LPCs, including 277 comets with  an original $1/a$ in the range [0, 100]\,au$_{-6}$;  
 numbers of comets as a function of their original $1/a$ are given in Table~\ref{tab:CODE_in_brief} whereas the overall distributions in $1/a_{\rm ori}$ and $q$ are shown in Fig.~\ref{fig:CODE_q_vs_aori}. 

\subsection{Why there are so many different orbits for some LPCs?}\label{sub:code_multiple_solutions}

It is a truism to state that, for each near-parabolic comet, it is worthwhile to determine an optimal orbital solution. However, implementing this idea is sometimes challenging, and a fully automated approach to orbit determination may overlook the highly individual characteristics of some comets. In the CODE catalogue, we chose to adopt a {\it preferred} solution based on the full observational data arc (using the NG~model of motion whenever possible). For comets that exhibit violent activity (such as outbursts or disintegration), we instead relied on the longest possible data arc not covering the extremal activity interval \citep{kroli-dyb:2020}. When NG~effects significantly influence the orbital motion, orbital solutions inevitably depend on additional assumptions about gas sublimation. In such cases, it is sometimes more reliable (e.g. for past evolution studies) to avoid introducing these assumptions and instead restrict the data arc used for orbit determination to  the pre-perihelion observations only \citep[see][and references therein]{Kroli-Dones:2023}.

More than 40\% of all LPCs in the CODE catalogue exhibit detectable trends in the [O--C] distribution over time. Among comets with small perihelion distances ($q < 3.35$\,au was assumed, see Section \ref{sec:CODE_a-distributions}), this behaviour is observed in more than 60\% of the cases. In the remaining comets, attempts to determine non-gravitational (NG) orbits did not yield results  that were reliable enough for inclusion in the CODE catalogue.

As a result of various possible force models and alternative treatments of the available positional data, the CODE catalogue presents multiple orbital solutions for well-observed comets. Among them, for each object, a \textit{preferred orbit} is indicated. For a detailed discussion of the NG~motion model used and numerous examples of  the NG trends, see \cite{kroli-dyb:2012, krolikowska:2020, kroli-dyb:2020, Kroli-Dones:2023}.

Thus, more than 40\% of all \textit{preferred orbits} in the catalogue are NG~orbits, for which the NG~parameters of the best-fitting force model are explicitly listed.
However, each NG~model introduces restrictive assumptions, particularly regarding the type and mechanism of ice sublimation from the cometary nucleus. One key assumption is that the NG~parameters remain constant throughout the  entire observational data arc \citep{marsden-sek-ye:1973,krolikowska:2020}.
Although this assumption is currently the only viable way to determine NG~orbits for a large number of LPCs, it is quite limiting \citep{Kroli-Dones:2023}.

It can be demonstrated that, for comets observed both long before and long after perihelion, it is possible to derive separate NG orbits that reveal different activity levels pre- and post-perihelion; see \cite{kroli-dyb:2012, Kroli-Dones:2023}.
Furthermore, when sufficiently rich positional data are available at large distances on a pre-perihelion leg, purely gravitational orbits fitted to it may, in fact, better represent the comet's past dynamical evolution and have the advantage of being free from additional assumptions about cometary activity.

Taking into account these factors, the CODE catalogue also marks some orbits as \textit{dedicated orbital solutions for backward dynamical evolution studies} (hereafter PB solutions), which may differ from the \textit{preferred orbits} where appropriate. In the next section, we present the original $1/a$ distributions derived from both sets of orbital solutions. Similarly, the catalogue also provides {\it dedicated orbits for forward evolution studies}.

\subsection{Why there are two different models for previous and next orbit calculation?}
\label{sub:code_two_models_prev_next}

As was briefly announced in \cite{Kroli-Dyb:2023} we changed  the way of presenting 'previous' and 'next' orbit elements, i.e. orbital parameters one revolution to the past and to the future. Including these orbits is a unique property of the CODE catalogue; therefore, we focus on the best way  to present them. Instead of showing one 'previous' and one 'next' orbit for each solution, we decided to offer the results of two different calculations: with and without  the stellar perturbation taken into account.

There are two main reasons for presenting two different sets of 'previous' and 'next' orbits. First, the results obtained from a pure Galactic potential would not change much in the future since our knowledge about the local Galactic potential seems quite satisfactory \citep[see][and references therein]{Dyb-Breiter:2021}. In  the case of stars, we expect to discover more stellar perturbers, especially multiple systems - work is in progress in this field \citep{Dyb_HD7977:2024}. As a result, the numbers resulting from the full dynamical model (Galaxy + stars) are subject to change, especially for smaller $1/a_{\rm prev}$. For now we are based on the latest {\it Gaia} mission results \citep{Gaia-DR3-release:2022,gaia-dr3-binaries:2022} but it is expected that the new results will be published soon.

 The second, and probably more important, aim of this change is to demonstrate how essential stellar perturbations are from the perspective of obtaining the previous perihelion distance and the resulting assessment of a comet's dynamical status. It is discussed in detail in Section \ref{prev_next}. 

\subsection{Changes in the CODE database interface}
\label{code_interface}

Several changes to the CODE database interface and content were already briefly announced in \cite{Kroli-Dyb:2023}. The most important  aspect is a method of presenting the 'previous' and 'next' orbits. In contrast to the previous versions, we now present them as two separate variants: resulting from calculations that take into account both Galactic and stellar perturbations and, for the sake of comparison, from calculations where stellar perturbations were completely omitted (see also  the previous section). Orbits of the second variant are named 'previous\_g' and 'next\_g'. The reasons for such a presentation are discussed in Section \ref{prev_next}. As the latest modification, we  also added the possibility of searching  for orbits separately among those obtained with and without stellar perturbations taken into account.

In  the case of comets with rich observational material, we offer several different orbital solutions, as described in Section \ref{sub:code_multiple_solutions}. Among them, there is always one solution  referred to as 'preferred'. In the current CODE database release, we additionally, for some comets, distinguish the particular solution as the \textit{best for comet past motion studies} (PB) or \textit{preferred for future motion analysis}. Currently, in the CODE interface, we clearly mark these 'special purpose' orbital solutions if they are different from the generally preferred orbits.  

We also extended the search functionality of the CODE interface, allowing for  a separated search among  the previous and next orbits in both variants, i.e. with and without stellar perturbations included in their calculations. As a result, the search page offers seven different orbits for each orbital solution: previous\_g, previous, original, osculating, future, next, and next\_g. In each case, it is also possible to restrict the search results to a preferred orbit. All documentation of the CODE database is updated and extended, and all changes are listed in the Changelog.

\section {Original $1/a$ distribution of LPCs from CODE catalogue.}\label{sec:CODE_a-distributions}

\begin{table*}[ht]
    \caption{Qualitative comparison between small-$q$ and large-$q$ samples of LPCs in the CODE catalogue; $1/a_{\rm ori}$ values in the second line are according to {\it preferred} solutions.}    \label{tab:CODE_qsmall_vs_qlarge}
\setlength{\tabcolsep}{6.0pt}    \centering
    \begin{tabular}{ccc}
	\hline \hline 
     &  small-$q$ sample ($q<3.35$\,au)  &  large-$q$ sample ($q\ge 3.35$\,au) \\
	\hline 
Number of LPCs                         &  187          &  182 \\    
Number of LPCs in the $1/a_{\rm ori}$-range of [0,100]\,au$_{-6}$ &  138          &  139  \\
LPC discoveries until 2000             &   83 (44\%)   &  37 (20\%)  \\
NG effects detectable                  &  often (over 60\%)  &  $\sim$3-times less often   \\
$1/a_{\rm ori}<0$                      &  12                 & 6 \\
	\hline  
    \end{tabular}
\end{table*}

\begin{figure*}
    \centering
    \includegraphics[width=0.45\linewidth]{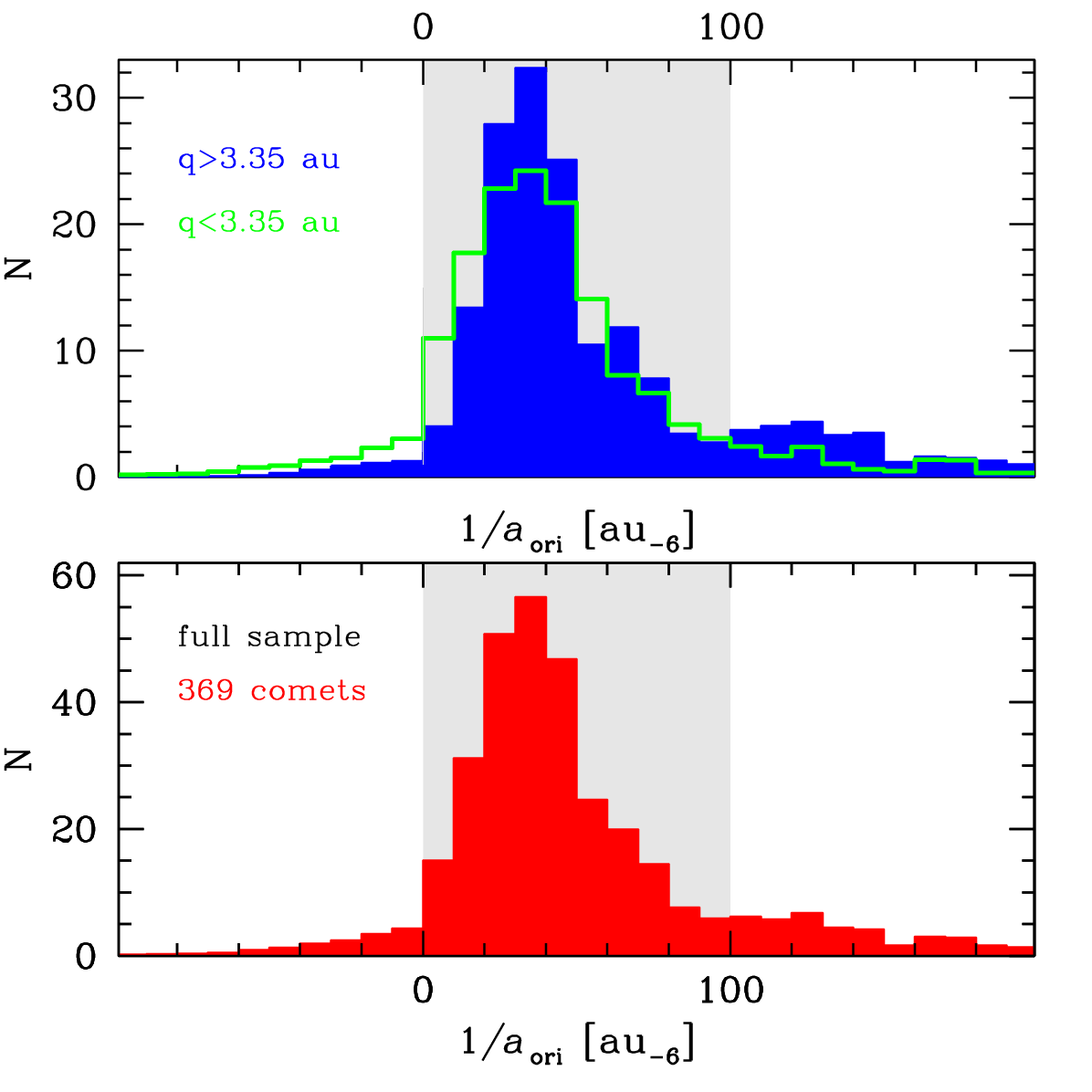}
    \includegraphics[width=0.45\linewidth]{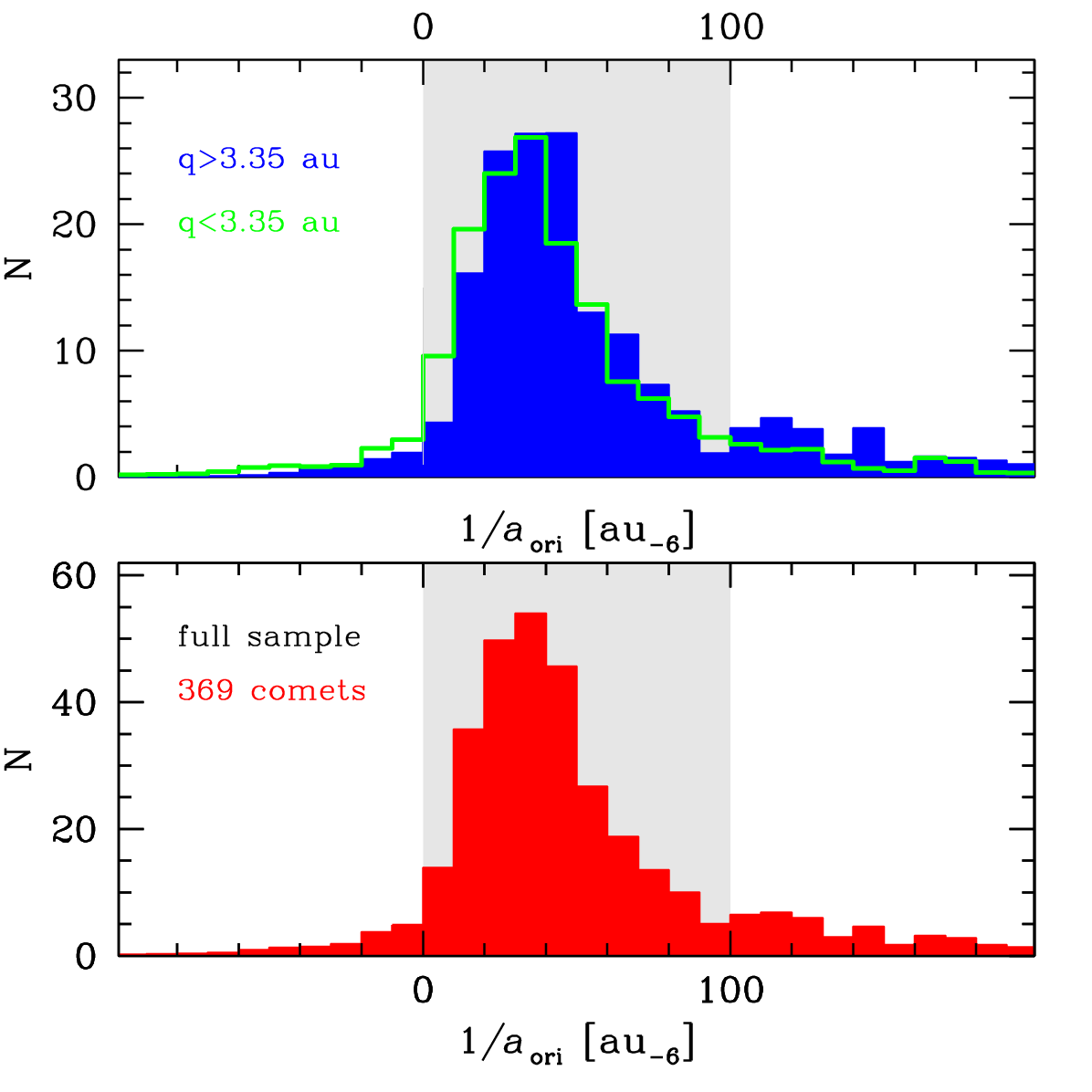}
    \caption{Distributions $1/a_{\rm ori}$ for two samples of LPCs:  small perihelion comets (green line histogram in the upper panels) and large perihelion comets (full blue histogram in the upper panels) and the sum of all LPCs included in the CODE catalogue (full red histogram in the lower panels); swarms of 5001\,VCs were used, bin width: 10\,au$_{-6}$. Left side panels show statistics using the 'preferred' orbits while the right side panels -- statistics based on PB solutions (i.e. orbits preferred for backward orbital evolution).}
    \label{fig:CODE_aori-distr}
\end{figure*}

\begin{table*}[ht]
    \caption{Deciles for original $1/a$ distributions for two sets of preferred orbits presented in the CODE catalogue, where only comets in the range [0,100]\,au$_{-6}$ for  original $1/a$ are taken for statistics.
    }    \label{tab:CODE_qsmall_vs_qlarge_deciles}
\setlength{\tabcolsep}{7.0pt}    \centering
    \begin{tabular}{ccccccccc}
	\hline \hline 
     &  \multicolumn{4}{c}{{\it preferred}  orbits}  &  \multicolumn{4}{c}{{\it orbits preferred for past evolution}}   \\
	\hline 
        & 10\%   & 50\%   & 90\%   & number &  10\%  & 50\%   &  90\% & number \\
small-$q$ & 14.16  & 36.89  & 71.01  & 138    &  14.04 & 34.43  & 70.73 & 138 \\
large-$q$ & 19.64  & 38.64  & 68.85  & 139    &  17.80 & 38.94  & 70.75 & 140 \\
all     & 16.44  & 38.2   & 70.77  & 277    &  15.79 & 37.15  & 70.80 & 278 \\
	\hline  
    \end{tabular}
\end{table*}

Here, we divided the entire sample of LPCs in the CODE catalogue into small and large perihelion groups based on a value close to the median value of $q$ for {\it preferred} set of orbits  in the range of [0,100]\,au$_{-6}$ for  the original $1/a$. 

According to this criterion, we chose the value of $q_{\rm lim}=3.35$\,au, which gives 138 and 139 Oort spike comets on the original bound orbits in  the small-$q$ and large-$q$ groups, respectively, while the full samples, including hyperbolas, consist of 182 and 187 comets in each group. Table~\ref{tab:CODE_qsmall_vs_qlarge} provides the comparative characteristics of both samples. 

The distributions of $1/a_{\rm ori}$ for LPCs with small perihelia ($q < 3.35$\,au) and large perihelia ($q \ge 3.35$\,au)  are shown in the upper left panel of Fig.~\ref{fig:CODE_aori-distr}, while the distribution for the entire LPC sample is presented in the lower left panel. Using the same threshold value of $q_{\rm lim}$, analogous statistics can be obtained for the PB orbital solutions. The results of this procedure are presented in the right panels of Fig.~\ref{fig:CODE_aori-distr}.

We note that all distributions shown in both panels are based on  the full swarms of 5001 VCs, which  ensures that orbital uncertainties are fully taken into account. The composite $1/a_{\rm ori}$ distribution is obtained by summing the individual $1/a_{\rm ori}$ distributions for all VCs of all comets in the considered samples, and divided by 5001 so that the vertical axis refers to the number of comets rather than the number of VCs; an example illustrating this procedure in detail, including the individual VC distributions, is presented in \cite{dyb-kroli:2016}, where these distributions were normalised to the number of clones instead of unity.

It should be noted that the distributions in the right panels -- based on the PB solutions -- are often based on shorter observational arcs (e.g.  the pre-perihelion leg of  the orbit). With respect to only this factor, the individual swarms of VCs in the right panels can be more dispersed than those shown in the left panel if only GR orbits were determined for a given comet. On the other hand, NG orbits used for the left panel distributions are often replaced by GR~orbits fitted solely to distant pre-perihelion data in the right panel distributions (see Sect.~\ref{sub:code_multiple_solutions}). Accordingly, from a qualitative perspective, the effect of orbital uncertainties should be similar in both panels.

The situation may be different when the small-$q$ and large-$q$ distributions within each panel are compared separately.  Due to the higher proportion of NG~orbits (see Table~\ref{tab:CODE_qsmall_vs_qlarge}) and typically shorter observational arcs, the small-$q$ distributions may exhibit greater scatter than the large-$q$ distributions; this effect is expected to influence both wings of the small-$q$ distribution. Keeping this in mind, we identify the following features in both upper panels of Fig.~\ref{fig:CODE_aori-distr}:

\begin{enumerate}
    \item For $1/a_{\rm ori} < 20$\,au$_{-6}$, small-perihelion Oort spike comets outnumber their large-perihelion counterparts by approximately 60\%.
    \item For $1/a_{\rm ori} > 50$\,au$_{-6}$, both distributions are generally similar in shape and number; however, in the range [100,150]\,au$_{-6}$, large perihelion LPCs outnumber small perihelion LPCs by a factor of two.
\end{enumerate}

In the middle part (three bins between 20 and 50\,au$_{-6}$) of both distributions in both upper panels, one can observe remarkable differences that highlight the level of uncertainty that still persists in this region of the $1/a_{\rm ori}$ distribution. However, the overall distributions in the lower panels are much more similar.  

The way in which the above-mentioned features are reflected at three selected points in the $1/a_{\rm ori}$ distribution (the tenth, 50th, and 90th percentiles) is presented in Table~\ref{tab:CODE_qsmall_vs_qlarge_deciles}. It shows that the PB set of orbits yields more similar 10\% and 90\% deciles (compared to the {\it preferred} orbit set), but also larger differences in medians. Furthermore, if we define the width of each distribution as the difference between these deciles, the small-$q$ distribution in the PB set is about 3.7\,au$_{-6}$ wider than the large-$q$ one. In contrast, for the {\it preferred} orbits, the difference in width between the small-$q$ and large-$q$ distributions reaches about 7.6\,au$^{-6}$.  

We would like to conclude these considerations by highlighting our recommendation that, for studies of the original $1/a$ distribution,  the PB orbital solutions are more appropriate.

\begin{figure}
    \centering
    \includegraphics[width=2.4in,angle=270]{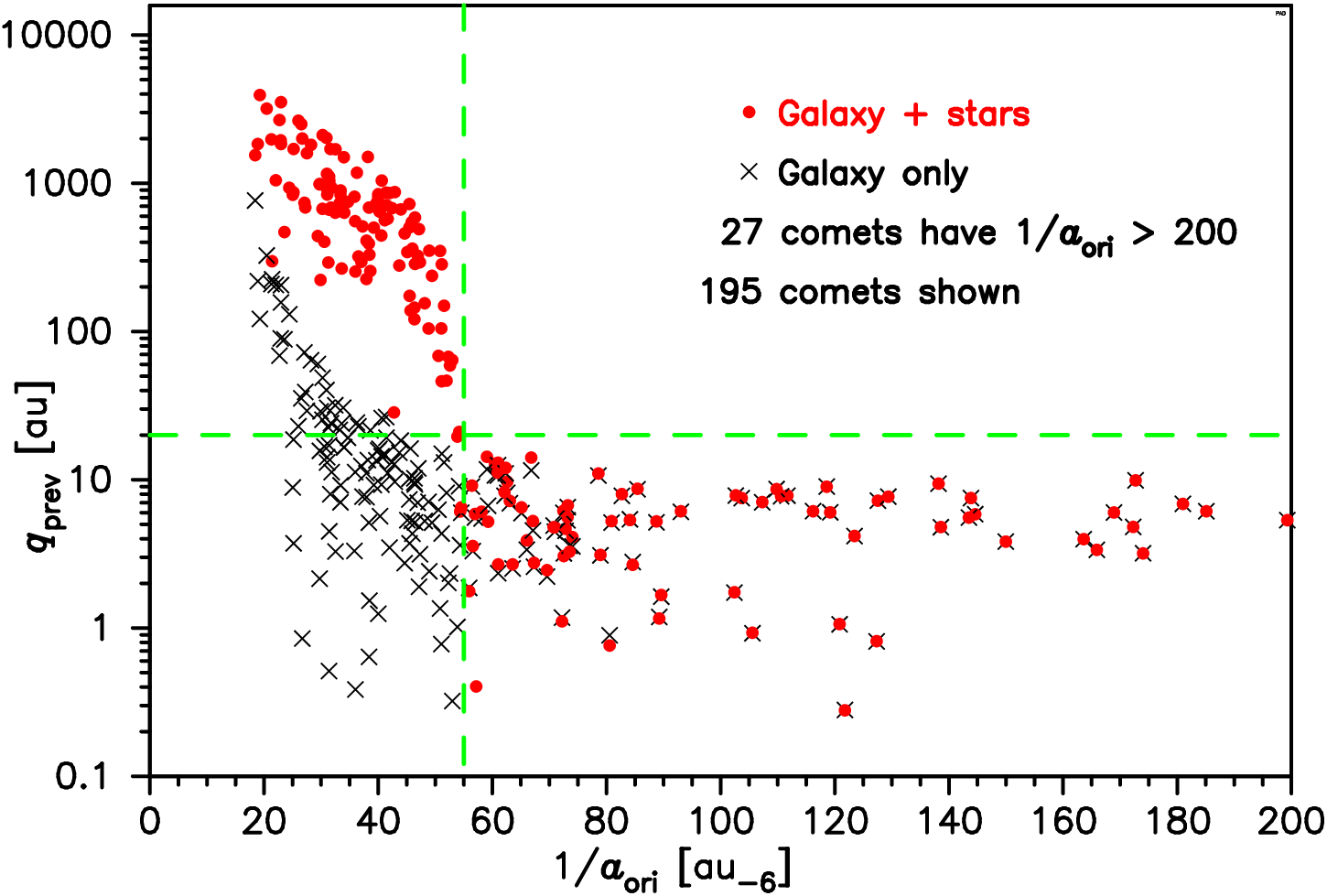}
    \caption{Previous perihelion distance dependence on an original semi-major axis reciprocal, only fully returning to previous perihelion  222} comets are used. Crosses mark the Galactic influence while red dots show both Galactic and stellar output. The horizontal dashed line depicts the assumed threshold for a dynamically new comet. The vertical dashed line divides the parameter space between an interval dominated by stellar and dominated by Galactic perturbations.
    \label{fig:q_prev_versus_a_ori}
\end{figure}

\section{Previous orbits and a comet dynamical status}
\label{prev_next}

The idea of accessing the dynamical status of LPCs on the basis of their previous perihelion distance (i.e., one orbital revolution to the past) was first proposed by \cite{kroli-dyb:2010}. The justification for such an approach is simple: if the previous perihelion distance ($q_{\rm prev}$) was small, a comet might have been significantly perturbed by planets, and it might also have experienced weaker or stronger heating from the Sun. We can call such a comet dynamically (and to some extent physically) old. Otherwise, if $q_{\rm prev}$ is sufficiently large, we call a comet dynamically new. After analysing many cases, we finally \citep{dyb-kroli:2015} adopted the following threshold values for $q_{\rm prev}$: a comet is dynamically new if this parameter is greater than 20~au. If it is smaller than 10~au we call a comet dynamically old, and for $10$~au $< q_{\rm prev} < 20$~au we describe a comet's status as uncertain.  To account for the orbital data uncertainties, we propagate to the past (and to the future) the nominal comet orbit and 5000 additional VCs. This approach has been used in CODE since the beginning, but now -- as described in Sect.~\ref{sub:code_two_models_prev_next} -- we present $q_{\rm prev}$ statistics in two variants resulting from the use of different dynamical models, with and without stellar perturbations.

In Fig.~\ref{fig:q_prev_versus_a_ori} we present the dependence of the previous perihelion distance on the original value of the reciprocal semi-major axis.  For both values we take a median of the corresponding VCs parameter distributions. Here, we compare  the $q_{\rm prev}$ values obtained from two different dynamical models (with and without stellar perturbations) as a function of $1/a_{\rm ori}$. All the results presented in this section are based on  the PB orbital solutions.

When constructing this figure, we only used 222 comets with a full swarm of VCs returning in both force models. This makes the statistics presented coherent and reliable; 27 comets with $1/a_{\rm ori}$ > 200 au$_{-6}$ are outside the frame, which makes the number of cases shown in Fig.\ref{fig:q_prev_versus_a_ori} equal to  195.

The vertical dashed line at 55\,au$_{-6}$ divides the interval of $1/a_{\rm ori}$ into two regions: to the right of it, the obtained $q_{\rm prev}$ values do not reflect stellar perturbations -- the results from the pure Galactic model ( black crosses) are almost identical to those from the full model ( red dots). In contrast, to the left of this line, in almost all cases, the inclusion of stellar perturbations results in significantly greater $q_{\rm prev}$ values.  

For all 222 fully returning comets,  the median of the previous perihelion distance obtained with stellar perturbations omitted equals 6.9~au, but after including stars it becomes equal  to 59.0~au. If we restrict this consideration to the left part of Fig.~\ref{fig:q_prev_versus_a_ori} (where $1/a_{\rm ori}$ < 55\,au$_{-6}$, 118 comets),  the median of $q_{\rm prev}$ with stars excluded (black  crosses) equals 12.7\,au, while from the full model (red dots) it equals 643.1\,au. From the point of view of a comet's dynamical status  and applying our 20\,au threshold  to these median $q_{\rm prev}$ values in the left part of the figure, we have  115 new comets,  none dynamically old and three uncertain cases (C/1890~F1, C/1941~K1, and C/2000~CT$_{54}$, the three lowest red points in this part) when stellar perturbations are taken into account. Without stars,  we obtain only 36 new comets with $1/a_{\rm ori}$ < 55\,au$_{-6}$.

Instead of  referring to a particular comet as dynamically old or new  on the basis of a median $q_{\rm prev}$ value,  in th CODE catalogue we present (for all its orbital solutions, not only for the PB ones) the statistics of the previous perihelion distance value. It is shown as the percentage of VCs having $q_{\rm prev}$ < 10~au, in the range of [10,20]~au or greater than 20~au.

 Apart from the 222 comets with fully returning VCs swarms and analysed above, we have 147 comets  with VCs swarms of 'previous' orbit partially or fully escaping. We set a threshold for the heliocentric distance of a comet equal to 120,000~au. The numerical integration of each VC motion into the past is stopped if a body  moves further than this  (and this VC is called escaping), otherwise it is stopped at the previous perihelion  (and called returning). Among these  147 comets  with mixed VCs swarms only 11 are fully escaping, with only two comets (C/1995~Y1 and C/2012~V1) having all VCs in hyperbolic orbits. For the remaining 136 comets, we obtained VC swarms in two parts: returning and escaping, with a subset of hyperbolas in the  latter. In these cases, we offer only partial statistics for the previous orbit parameters (see the CODE database documentation for more details). When stellar perturbations are included, 75 comets from this group of 136 comets have more than 80\% VCs returning; without stars, the same percentage is for 71 comets. Therefore, the statistics of $q_{\rm prev}$ and the resulting dynamical status assessments for these more than 70 comets are also quite reliable. There are only 21 comets with less than 50\% of  the VCs returning.

All of the above analysis leads to the conclusion that the widely used method of discriminating between dynamically old and new comets solely on the basis of the value of $1/a_{\rm ori}$ often  results in incorrect outcomes, especially with the threshold value of $1/a_{\rm ori} = 100$ au$_{-6}$. If one cannot calculate $q_{\rm prev}$, which we offer in the CODE catalogue, the better approximate rule for a dynamically new comet is $1/a_{\rm ori} < 55$ au$_{-6}$ (i.e. $a_{\rm ori} \gtrsim $ 20,000~au) assuming that we take into account stellar perturbations. If only Galactic perturbations are used, the discrimination between old and new comets on the basis of only $1/a_{\rm ori}$ seems { problematic; see also Fig.~13 in \cite{kroli_dyb:2017}}.

\begin{figure}
    \centering
    \includegraphics[width=1.03\linewidth]{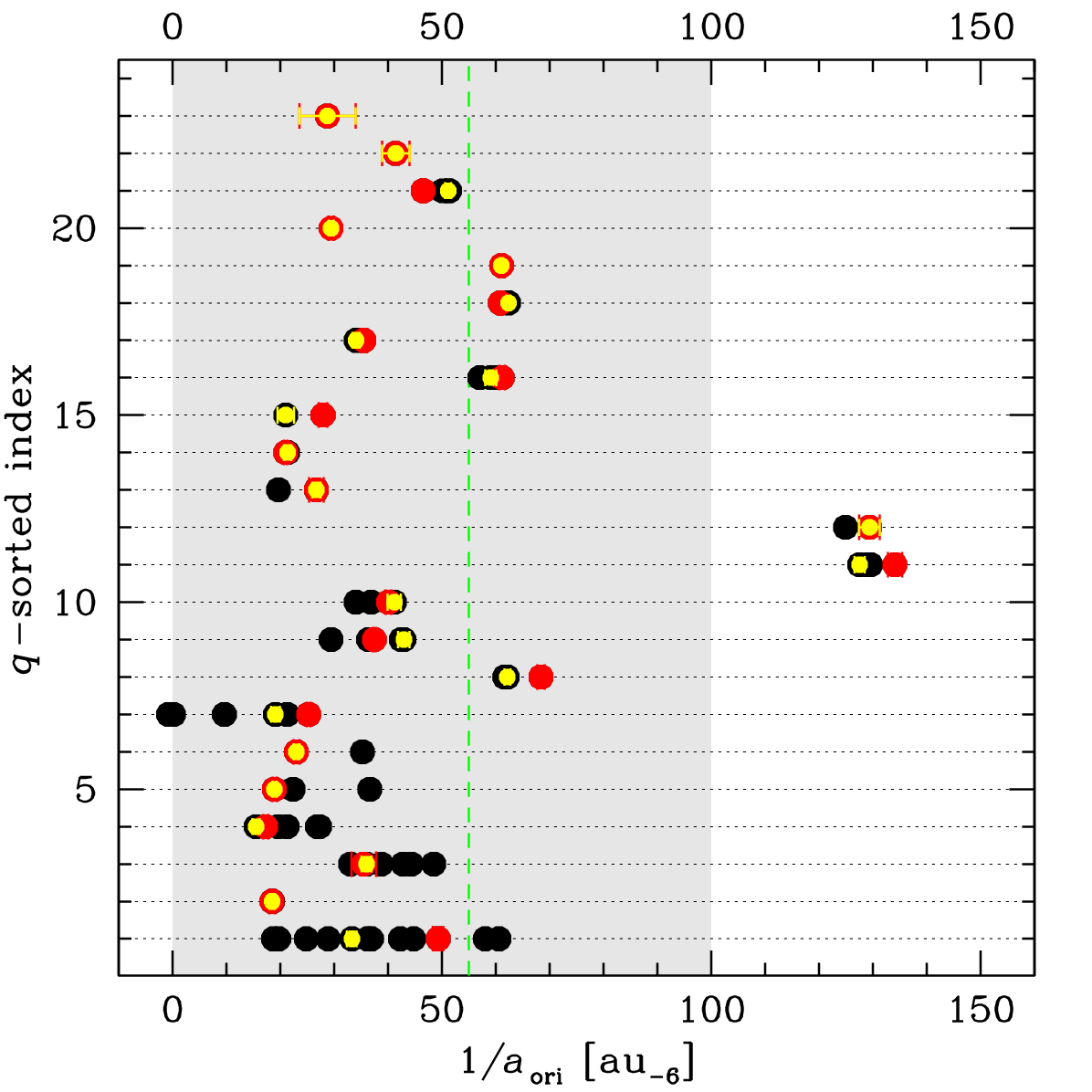}
    \caption{Values of $1/a_{\rm ori}$ for all solutions provided in the CODE catalogue for LPCs listed in Table~\ref{tab:Oortspikecomets10au}, excluding those based solely on the post-perihelion part of orbits. The vertical axis represents the ordinal number of comet when sorted by $q$. The same number is in the final column of Table~\ref{tab:Oortspikecomets10au}.  The {\it preferred} solutions with their uncertainties are shown in red, the PB solutions are marked by yellow points, and all remaining values are plotted in black. The green dashed vertical line indicates the same reference value as in Fig.~\ref{fig:q_prev_versus_a_ori}.} 
    \label{fig:Oortspikecomets10au}
\end{figure}

\begin{figure}
    \centering
    \includegraphics[width=1.00\linewidth]{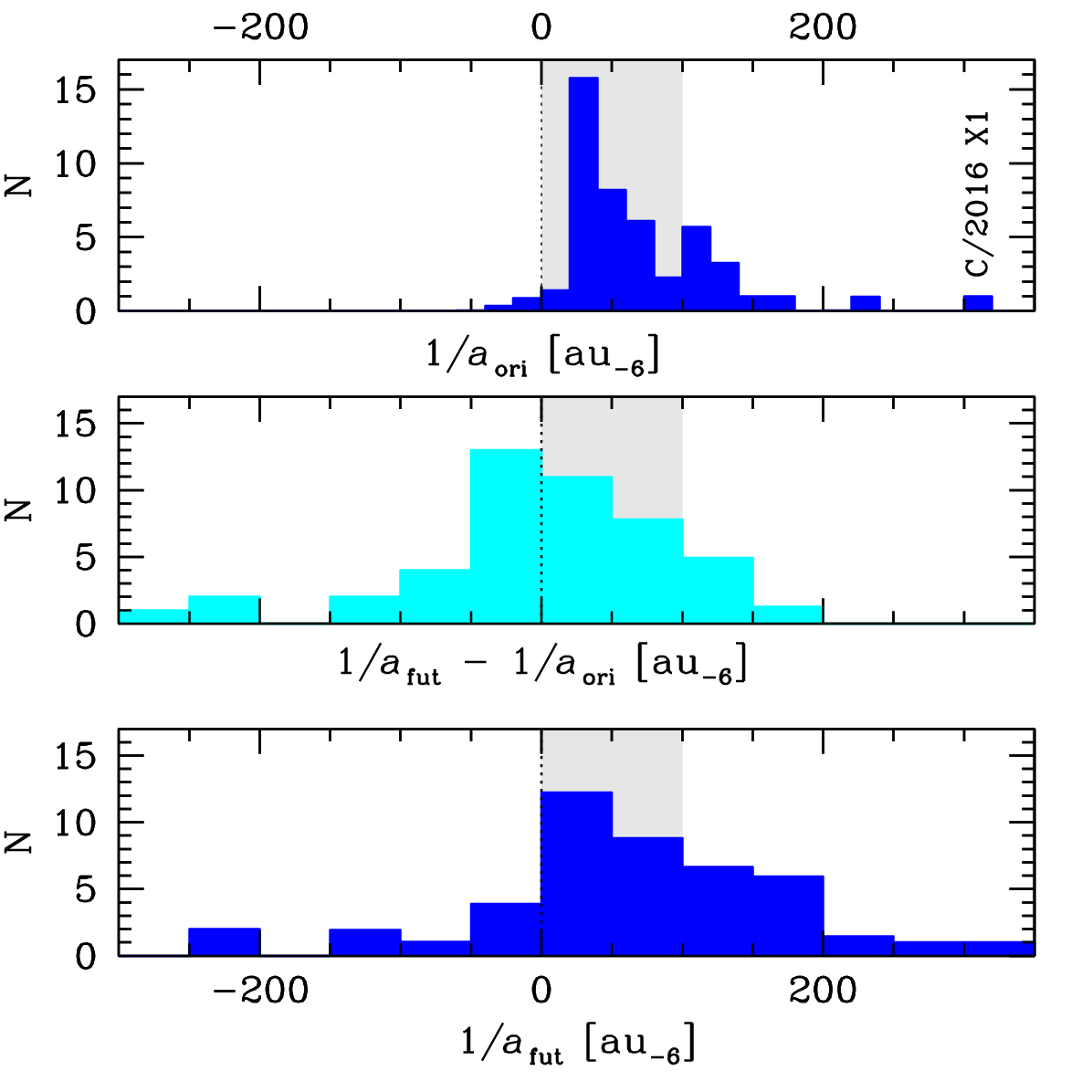}
    \caption{Sample of 48 LPCs from the CODE catalogue having $q>7$\,au. The upper and lowest [panels shows original and future distribution of 1/a, respectively while the middle panel gives the distribution of planetary perturbations during the passage through the planetary system. The light grey vertical band indicates the region occupied by Oort spike comets. The range of the horizontal axes is chosen to include the tails of the $a_{\rm ori}$ distribution.}
    \label{fig:CODE_qgt7-distr}
\end{figure}

\begin{table*}
\caption{\label{tab:Oortspikecomets10au} Complete list of LPCs discovered before 2022 more than 10~au from the Sun or with pre-discovery detections beyond 10~au included in the CODE catalogue. } \setlength{\tabcolsep}{4.0pt} \centering
	\begin{tabular}{llrccccccr}
	\hline \hline 
Comet      & $q$  & \multicolumn{2}{c}{heliocentric distance} & $T_{\rm per}$ &  $1/a_{\bf ori}$ & \multicolumn{2}{c}{dynamical status} & model & $q$-sorted \\
&& discovery   & pre-discovery &&& Gal & Gal+Stars & type & index \\
& [au] &  [au] & [au] & [yyyy\,mm\,dd]    & [au]$_{-6}$ & \multicolumn{2}{c}{}  & & \\
  & [2] & [3] & [4] & [5] & [6] & [7] & [8] & [9] & [10] \\
\hline
C/2001 Q4  & 0.96 & 10.1  & --   & 2004\,05\,15 & 33.3$\pm$0.9                       & DN71 & DN & NG & 1 \\
C/2003 A2  & \hspace{-0.2cm}11.4  & 11.5  & 11.9 & 2003\,11\,06 & 41.4$\pm$2.5                       & DN99 & DN & GR & 22\\
C/2005 L3  & 5.59 & 8.68\hspace{-0.2cm}  & 10.3 & 2008\,01\,16 & 62.1$\pm$0.6        & DO & DO & GR & 8 \\
C/2006 S3  & 5.13 & 13.3  & 26.1 & 2012\,04\,12 &19.1$\pm$0.8                        & DN & DN & NG & 7 \\
C/2008 S3  & 8.02 &  9.46\hspace{-0.20cm} & 12.4 & 2011\,06\,10 & 21.3$\pm$0.7       & DN & DN & GR & 14 \\
C/2010 U3  & 8.45 & 18.4  & 25.8 & 2019\,02\,26 & 59.1$\pm$1.0                       & DU & DU & GR & 16 \\
C/2012 LP$_{26}$  & 6.54  & 9.19\hspace{-0.2cm} & 10.0 & 2015\,08\,16 & 41.2$\pm$1.2 & DN & DN & GR & 10 \\
C/2014 B1  & 9.56 & 12.0  & --   & 2017\,09\,08 & 61.1$\pm$0.5                       & DO & DO & GR & 19 \\
C/2014 UN$_{271}$ & \hspace{-0.2cm}10.9  & 29.0 & 34.1 & 2031\,01\,22 & 51.1$\pm$0.6 & DU & DN & GR & 21 \\
C/2015 D3  & 8.15 &  7.69\hspace{-0.2cm} & 10.5 & 2016\,05\,03 &21.0$\pm$1.6         & DN & DN & GR & 15 \\
C/2016 Q2  & 7.08 & 12.6  & --   & 2021\,09\,28 & \hspace{-0.2cm}127.6$\pm$0.9       & DO & DO & GR & 11 \\
C/2017 E3  & 5.92 &  5.95\hspace{-0.2cm} & 10.3 & 2017\,05\,31 & 42.9$\pm$1.0        & DO75 & DN & GR & 9 \\
C/2017 K2  & 1.81 & 16.0  & 23.7 & 2022\,12\,19 & 36.0$\pm$0.9                       & DU86 & DN & NG & 3 \\
C/2019 E3  &\hspace{-0.21cm}10.3  & 12.9  &\hspace{+0.21cm}19.1& 2023\,11\,15 & 29.4$\pm$0.5& DN & DN & GR & 20 \\
C/2019 U5  & 3.62 & 10.4  & 10.5 & 2023\,03\,29 & 18.9$\pm$0.4                       & DN & DN & NG & 5 \\
C/2020 F2  & 8.82 & 10.1  & 11.0 & 2022\,07\,15 & 62.4$\pm$0.4                       & DU & DU & GR & 18 \\
C/2020 V2  & 2.23 & 8.71\hspace{-0.2cm}  & 10.1 & 2023\,05\,08 & 15.5$\pm$0.7        & DN & DN & NG & 4 \\
C/2021 A9  & 7.76 & 10.0  & 10.1 & 2023\,12\,01 & 26.7$\pm$0.8                       & DO & DN & GR & 13 \\
C/2021 G2 & 4.98 & 10.1  & 11.0 & 2024\,09\,09 & 35.2$\pm$0.2                       & DN & DN & NG & 6 \\
C/2021 Q4  & 7.56 & 8.58\hspace{-0.2cm}  & 10.2 & 2023\,06\,10 & \hspace{-0.2cm}129.4$\pm$1.9  & DO & DO & NG & 12 \\
C/2021 Q6  & 8.72 & 10.3  & 11.0 & 2024\,03\,21 & 34.1$\pm$0.7                       & DU & DN & GR & 17 \\
C/2021 S3 & 1.32 &  8.92\hspace{-0.2cm} & 11.0 & 2024\,02\,14 & 18.5$\pm$0.5         & DN & DN & NG  & 2 \\   \\
C/2025 D1 &\hspace{-0.21cm}14.1  & 15.1  & 21.3& 2028\,05\,19 & 28.7$\pm$5.2        & DN75 & DN & GR & 23 \\   
\hline
\end{tabular}
\tablefoot{
This is an updated and enriched version of Table~1 from \cite{Kroli-Dones:2023} (with new columns [6]--[10]). We list here the comets' perihelion distances ($q$); their heliocentric distances at discovery and in their earliest pre-discovery detection; and their dates of perihelion passage $T_{\rm per}$. 
Values for $1/a_{\rm ori}$ (column [6]), and dynamical status estimates (columns [7]-[8]) are given for the PB solutions in the CODE catalogue. Dynamical status  is given according to the statistics of 5001 VCs $q_{\rm prev}$ values. 
DN, DU, DO mean the dynamically new, dynamically uncertain, and dynamically old comet, respectively, and a two digit number given in a few cases after the status mean the percentage of VCs having the declared status. The last column gives an ordinal number of comet when sorted by $q$.
}
\end{table*}

\section{Comets discovered beyond 10\,au and their dynamical status} \label{sec:distant_discoveries}

Due to the increasing technical capabilities, for example those currently realised by the Vera Rubin Telescope, there is a growing interest in comets discovered from distances beyond the orbit of Saturn. In Table~\ref{tab:Oortspikecomets10au} we provide a list of 23 such comets. Only four have $q$ below 3.35\,au, and 17 of these 23 have $1/a_{\rm ori}$ ranging from 10 to 55\,au$_{-6}$ (the threshold value discussed in Sect.~\ref{prev_next}), while the remaining six comets have original semi-major axes shorter than 20,000\,au (see column [6]). Columns [7]-[8] show the dynamical status of these LPCs in two dynamical models: with and without stellar perturbations; see \cite{kroli_dyb:2017} and Section \ref{prev_next} for a discussion.  

One can see that 12 of these 23 comets seem to be dynamically new when only  the Galactic tide is used; however, with  the stellar perturbations included, the number of dynamically new comets grows to 17. We have only four dynamically old comets here, two with $1/a_{\rm ori}$ > 100\,au$_{-6}$ (C/2016~Q2 and C/2021~Q4) and two having the original $1/a$ between 60-70\,au$_{-6}$ (C/2005~L3 and C/2014~B1). 
There are also two comets in both models classified as 'uncertain'. They also have $1/a_{\rm ori}$ between 60 and 70\,au$_{-6}$, but the previous perihelion distance for all their VCs lies between 10\,au and 20\,au  (C/2010~U3 and C/2020~F2). More uncertain and dynamically old comets are in column [7].

It is obvious that these estimates of dynamical status depend on our knowledge of the passing stars and may be subject to modification in the future for an individual object. However, the underlying statistical message that most near-parabolic comets detected so far at distances beyond 10\,au are dynamically new does not appear to be  changing.

For most comets listed in Table \ref{tab:Oortspikecomets10au}, we also present several different orbital solutions, as explained in Section \ref{sub:code_multiple_solutions}. Although these are comets with large perihelion distances, reviewing the solutions in the CODE catalogue reveals that NG~effects are evident in the motion of at least eight of them  (column~[9]). Six of these comets (C/2005~L3, C/2006~S3, C/2008~S3, C/2010~U3, C/2012~LP26, and C/2015~D3) are  the subject of a detailed discussion in \cite{Kroli-Dones:2023}, where it is shown that the motion of  the comets is often measurably affected by NG~forces at heliocentric distances further than 5\,au from the Sun. The most spectacular example is C/2010~U3 (Boattini) with  a perihelion distance of 8.45\,au. 

The CODE~Catalogue is the only existing  catalogue that allows the user to investigate how different approaches to selecting the data arc for a model of motion (GR or NG) affect both the values of $1/a_{\rm ori}$ and $q_{\rm prev}$. These values are crucial to assessing whether the comet studied is dynamically old or new. To assess the reliability of the conclusions drawn from Table~\ref{tab:Oortspikecomets10au}, which is based on a single orbital solution (the orbits preferred for the past dynamics studies, PB solutions) Fig.~\ref{fig:Oortspikecomets10au} will be helpful. 

This figure shows all available orbital solutions provided in the CODE catalogue, where the {\it preferred} orbits are shown by red dots, and PB solutions are indicated by smaller yellow dots. One can see that both types of orbits are different for 11 LPCs; however, only for  the C/2001~Q4 (the first case with the smallest value of $q$)  differs markedly. For the PB solution used in Table~\ref{tab:Oortspikecomets10au} this comet is classified as dynamically new in both models (slightly weaker when stellar perturbations are omitted). But for the preferred orbit, it is dynamically old without stars. For the rest of  the comet's orbital solutions, the situation varies remarkably; see also \cite{kroli-dyb:2012}.  

\section {LPCs with perihelion distance greater than 7\,au}\label{sec:CODE_large_q}

Fig.~\ref{fig:CODE_q_vs_aori} shows that the number of LPCs with perihelia beyond Jupiter's orbit systematically decreases due to current observational limitations.  
Here we describe the LPC sample with $q$ above 7\,au. The only such distant comet known before 2000 is C/1999~J2 (Skiff), with $q = 7.11$\,au.  
Currently, we  have found 56 such LPCs using the JPL Small Body Database Query \citep[][as of May 2025]{JPL_SBDB}. The JPL list includes 45~LPCs that are also present in the CODE catalogue, among which twelve are outside the Oort spike.  
It should be noted that  the three Oort spike comets present in the CODE and having $q > 7$\,au are absent from the list of 56 comets mentioned above because they are classified in JPL as  Chiron-type comets (C/2016~X1, $1/a_{\rm ori} = 310.1 \pm 1.2$\,au$_{-6}$ and C/2017~AB5, $1/a_{\rm ori} = 66.82 \pm 2.33$\,au$_{-6}$) or as  JFCs (C/2007~D1, $1/a_{\rm ori} = 43.96 \pm 0.95$\,au$_{-6}$).  

Two of the twelve not included in CODE have  an original $1/a$ in the range  of 100--200\,au$_{-6}$; the remaining ones have shorter semi-major axes.  
Together, this means that only about 20\% of the known LPCs with $q > 7$~au have semi-major axes shorter than 5,000\,au. The smallest semi-major axis belongs to C/2024~E2 (Bok), with $a = 51$\,au and an orbital period of 364 years.  

Fig.~\ref{fig:CODE_qgt7-distr} shows the distributions of the original $1/a$ values for all LPCs with $q > 7$~au and present in the CODE catalogue. We used the full swarm of 5001\,VCs for each comet to account for the orbit uncertainties, as in Fig.~\ref{fig:CODE_aori-distr}, and here we use the {\it preferred} orbits to be consistent in the values of $1/a_{\rm fut} - 1/a_{\rm ori}$. In the lowest plot, two LPCs are outside the right border: C/2016~X1 with $1/a_{\rm fut} = 503.8 \pm 1.2$\,au$_{-6}$ and C/2007~D1 with $1/a_{\rm fut} = 738.4 \pm 1.0 $\,au$_{-6}$. It appears that almost all of these LPCs have original $a$ below 50,000\,au, and the peak between 25,000\,au and 50,000\,au includes 34\% of these comets (highest bin in the upper plot of Fig.~\ref{fig:CODE_qgt7-distr}). 

Moving to shorter original semi-major axes, we observe a numerous group of comets between $\sim$7,000\,au and 25,000\,au, comprising about 54\% of the sample given in the CODE catalogue, as well as a more dispersed population with original semi-major axes shorter than 7,000\,au. It is worth noting that moderate planetary perturbations (middle panel in Fig.~\ref{fig:CODE_qgt7-distr}) caused only 9 (<20\%) of these comets to leave the inner part of the Solar System on hyperbolic orbits in the future.

\section{Prospects and conclusions}

For obvious reasons, maintaining a catalogue/database such as CODE is a never-ending story. Each year, we discover more and more comets; we observe them at larger distances and over longer time spans. Our individualised approach and careful model fitting make the orbit determination time-consuming. Additionally, for a number of comets already discovered, we have to wait for more observations. All of this means that the CODE database is almost complete for older comets, but many comets discovered in 2022 and later will be included in the future.

The CODE catalogue offers the results of  a long-term propagation of comet motion to the previous and next perihelion. This is necessary to discuss  the the comet's dynamical status as well as the overall influence of planetary perturbations on the entire population of long-period comets. To achieve this, it is necessary to take into account perturbations caused by stars passing near the Sun during the comet's previous or next orbit and the overall Galactic potential. To this  end, we used modern, fast, and precise algorithms and data proposed by \cite{Dyb-Breiter:2021}. Currently, we are based on a list of potential stellar perturbers taken from release 3.3 of the StePPeD\footnote{\tiny{\tt https://apollo.astro.amu.edu.pl/StePPeD }} database. Work is in progress to find more perturbers \citep{Dyb_HD7977:2024}, especially among multiple systems, so more or less significant changes to the previous and next orbits are expected in the future.

Our conclusion is that such an individualised approach to each comet, careful model fitting, various strategies  for the selection of observational material, and the preparation of different, dedicated orbital solutions are our valuable contributions to the studies of LPCs' source regions and origins. The message to the reader is quite straightforward: If one wants to study the origin of LPCs (or, for example, the structure of the Oort cloud), one should restrict oneself to dynamically new comets. To discriminate between dynamically old and new comets, it is not sufficient to calculate $1/a_{\rm ori}$;  rather, $q_{\rm prev}$ is necessary.  However, in calculating $q_{\rm prev}$, stellar perturbations might be essential. In the updated CODE catalogue, we offer all these parameters based on the best and current data. 

\paragraph{{\bf Data availability}}

The data underlying this article are available at {\tt https://apollo.astro.amu.edu.pl/CODE} or via {\tt https://code.cbk.waw.pl/}.  Additional material will be shared  upon reasonable request to the corresponding author.

\begin{acknowledgements}
The CODE Catalogue has made use of  the positional data of comets provided by the International Astronomical Union’s Minor Planet Center. We acknowledge the use of some data downloaded from \textit{Solar System Dynamics}: https://ssd.jpl.nasa.gov as stated in the main text.
 We would like to thank the anonymous referee for constructive remarks.
\end{acknowledgements}

\bibliographystyle{aa}   
\bibliography{PAD32} 

\end{document}